# Optimal Phasor Measurement Unit Placement Using a Honey Bee Mating Optimization (HBMO) Technique Considering Measurement Loss and Line Outages

Lorena Palmero[1], Ugur Saritac[2*], Ali Dehghan Chaharabi[3]

[1] *Spanish Institute for Foreign Trade (ICEX España Exportación e Inversiones), Spain*
[2] *Department of Electrical Engineering, Lamar University, Beaumont, TX, U.S.A*
[3] *Department of Electrical Engineering, Islamic Azad University, Iran*

**Abstract**

Phasor measurement units (PMUs) are important devices for protection, monitoring, and control of modern power systems. Unlike the supervisory control and data acquisition (SCADA) system which only measure the magnitude, PMUs can provide a set of synchronized phasor measurements with high sampling rate with the accuracy of less than one second, for better system monitoring especially during the outages. Since the PMUs are costly devices, system operator is not able to install a PMU at each bus to maximize the observability of system. To this end, determination of the number of PMUs in the system for having a 100% observability is pivotal. In this paper, honey bee mating optimization (HBMO) approach is utilized to allocate PMUs in modern power systems. Moreover, the impact of line outages, as well as measurement loses, are considered. The simulation results show the effectiveness of the proposed method on solving the PMU placement problem.

## 1. Introduction

State estimation is a pivotal process for improving the power system monitoring, control, and security. Remote terminal units (RTUs) were responsible for collecting measurements like real/reactive power flows, power injections, and magnitude of bus voltages and branch currents for supervisory control and data acquisition (SCADA) as the source provider of state estimation. In this system, the measured data is not precise enough for system studies due to the lack of simultaneously measured and transferred data in long distances. Moreover, inaccessibility to the phase angle of busses' measured voltages and currents is another disadvantage of this system. In the 1990s, phasor measurement units (PMUs) were introduced for being utilized in wide area monitoring of systems (WAMS) [1, 2]. Nowadays, PMUs are considered as critical devices for system operators to monitor and control the system. In order to explain the observability concept, if a PMU is installed on bus $i$ then bus $i$ and all adjacent buses connected to bus $i$ are observable. It means that the voltage and current phasors of these buses are known. The system is called observable if every bus properties of the network is observable. The PMU placement concept is simply explained as follows: If a bus or at least one of its adjacent buses is equipped with PMU then this bus is observable. The best solution to make a system fully observable is to install a PMU device per bus. However, due to the higher cost of these devices, this is not a cost effective means of PMU installation. As a result, based on the budget of system operator, the best locations must be selected for installing the PMU devices.

There are several research studies with different heuristic algorithms focused on optimal PMU placement challenge. Genetic algorithm (GA) is one of the nature-based algorithms and can be used for different purposes of modeling the power systems [3]. Focusing on PMU placement problem, in [4], a GA was used for determination of the number of PMUs in system. In this method the zero-injection bus (ZIB) model was considered to maximize the observability based on ZIB buses. Authors of [5] provided a solution for optimal number of PMUs in power systems using GA technique. Maximum redundancy as well as branch and PMU loss were considered in this paper. Another well-known heuristic algorithm is the particle swarm optimization (PSO) method which widely utilized in power systems. In PMU placement problem, authors of [6] proposed a novel concept for solving the PMU placement problem which focused on transmission line outage and measurement losses based on binary PSO algorithm. Additionally, in [7], the PSO algorithm was adapted to address the observability maximization of PMUs considering configurability. The authors also used the fault-tolerant PMU installation problem as a constraint to minimize the total number of PMUs. Differential evolution (DE) algorithm is another heuristic concept that utilized in PMU allocation problem. Authors of [8] analyzed the network observability based on DE algorithm. The simulation results were presented based on ZIB buses. In [9], optimal placement of PMUs was discussed based on mean square error minimization using DE algorithm.

* Corresponding Author: Ugur Saritac
E-mail address: usaritac@lamar.edu

Matrix reduction technique is another effective method for solving the power system related challenges. Considering the PMU placement problem, in [10], a matrix reduction algorithm was used to determine the number of PMUs. It should be noted that this method was combined with greedy search algorithm for better determination the number of PMUs. Besides matrix reduction algorithm, a new integer linear programming method is applied to PMU placement problem in [11] and [12] for minimizing the cost of installation for robust static state estimation. A teaching learning-based optimization technique was proposed in [13] by considering different objective functions including total number of PMUs, observability, and voltage stability constraints. The proposed framework addressed the PMU allocation problem considering ZIBs, line outage and PMU loss. In addition to the above-mentioned algorithms, PMU placement problem was also investigated based on cellular learning automata (CLA) [14] with the goal of minimizing the total cost of PMU installments considering the measurement redundancy. In [15], a mixed integer programming method was proposed to place PMUs considering total cost and measurement channels.

Considering the literature review, several methods were presented to optimally place the PMUs in power systems considering different objective functions. In this paper, a new method, known as honey bee mating optimization (HBMO) is applied to optimal PMU placement problem to effectively place PMUs in the system. ZIB buses, line outages, and PMU loss is also modeled to present the variety of possible scenarios.

The following sections will provide information regarding the 1) mathematical formulation 2) solution technique 3) simulation results, and 4) conclusion and future works.

## 2. Problem Formulation

In this section the PMU placement problem mathematical formulation with the aim of minimizing the total number of PMUs (total cost of installing PMUs) will be provided.

The main objective in OPP problem is to determine the minimum number of PMUs and their appropriate placements to ensure the power system observability (the constraint of the problem) discussed in the first section. The mathematical formulation of this part is presented as Eqs. (1) & (2):

$$\min \sum_{j=1}^{N} U_j \times c_j \tag{1}$$

$$OBSV_i = \sum_{j=1}^{N} U_j . K_{ij} \geq 1 \tag{2}$$

where $U_j$ and $c_j$ are the number of PMUs and cost of installing 1 PMU in power system, respectively [16]. Additionally, $OBSV_i$ and $N$ denote observability of system which must be greater than 1 to observe every bus with a PMU device, and total number of buses. Moreover, the connectivity matrix, shown as $K_{ij}$, show the connectivity of matrix and gets the value of 1 when the bus $i$ and bus $j$ were connected. Otherwise, it gets a value of zero.

To consider the impact of ZIBs in the optimization, following equations must be used to address the observability by adjacent buses.

$$OBSV_i = \sum_{j=1}^{N} U_j . K_{ij} + \sum_{j=1}^{N} z_j . K_{ij} . y_{ij} \geq 1 \tag{3}$$

$$z_j = \sum_{j=1}^{N} K_{ij} . y_{ij} \tag{4}$$

where $z_j$ and $y_{ij}$ are binary parameter and auxiliary parameter to model the ZIBs. With the help of these equations, there is no need to install PMU units all over the system, because some of the buses can be observable based on ZIBs.

To model the possible contingencies such as PMU loss, following constraints must be considered:

$$OBSV_i + z_j \geq MOB \tag{5}$$

$$z_j = \sum_{j=1}^{N} K_{ij} . y_{ij} \tag{6}$$

where $MOB$ denotes minimum observable buses, which should be considered as Eq. (2). The reason behind that is, in the case of any disruption, or attack to one PMU, at least another PMU monitors the bus.

In addition to the measurement loss, the system may experience line outages scenarios which highly affect the observability of entire system. To that end, following constraints must be prepared for any line outages in transmission system.

$$OBSV_i^l = \sum_{j=1}^{N} U_j . K_{ij}^l + \sum_{j=1}^{N} z_j . K_{ij}^l . y_{ij}^k \geq 1 \tag{7}$$

$$z_j = \sum_{j=1}^{N} K_{ij}^l . y_{ij}^k \tag{8}$$

Considering the above-mentioned equations, $l$ denotes to the line number. Moreover, $K_{ij}^l$ is the connectivity parameter and when the line $l$ connects bus $i$ and bus $j$, get the value of 1, and in the case of line outages it gets the value of zero (which shows that the observed line is being out of service).

## 3. Solution Technique

For the first time, the HBMO algorithm was presented in [17-20]. These algorithms are based on GA and simulated annealing method with better efficiency compared to both algorithms. More details with great examples can be found in [21]. The algorithm starts with defining the algorithm parameters and model parameters. Then, a set of initial solutions will be generated (these initial solutions may or may not feasible solutions). Based on penalizing the objective function, the best predefined solutions are selected. In the next step, based on stimulated annealing, a set of solutions are selected for possible information exchange. Then, a new set of solutions are generated by using a predefined crossover heuristic functions and the results are getting improved. Finally, if the new best solution was better than previous solution then the HBMO algorithm finishes. More information regarding the HBMO algorithms is provided in Figure 1.

## 4. Simulation Results and Analysis

The effectiveness of the proposed HBMO based PMU placement problem is tested on IEEE 14, 57, and 118 bus test systems. The results for optimal placement of PMUs considering line outage, PMU loss, ZIBs together with HBMO algorithm will be presented in the following sections. Additionally, the IEEE test systems are provided for better understanding the location of PMUs. More information regarding the tests systems can be found in [22]. As mentioned above, ZIBs are considered as important buses for making the system observable. To this end, the number of ZIBs and their location are provided in Table 1.

**Table 1.** Number and Location of Zero-Injection-Buses

| Test Network | Number of ZIB | PMUs Location |
|---|---|---|
| IEEE 14 bus | 1 | 7 |
| IEEE 57 bus | 15 | 4, 7, 11, 21, 22, 24, 26, 34, 36, 37, 39, 40, 45, 46, 48 |
| IEEE 118 bus | 10 | 5, 9, 30, 37, 38, 63, 64, 68, 71, 81 |

### 4.1. Optimal PMU Placement Results with and without ZIBs

The results of PMU placement problem with and without considering the ZIBs are provided in Table 2 and Table 3. The simulation results show that with including the ZIBs in the optimization process, the total number of PMUs is reduced due to the observably opportunity that ZIBs can provide for the system. Additionally, the location of PMUs to make system 100\% observable is provided for both with and without considering ZIBs.

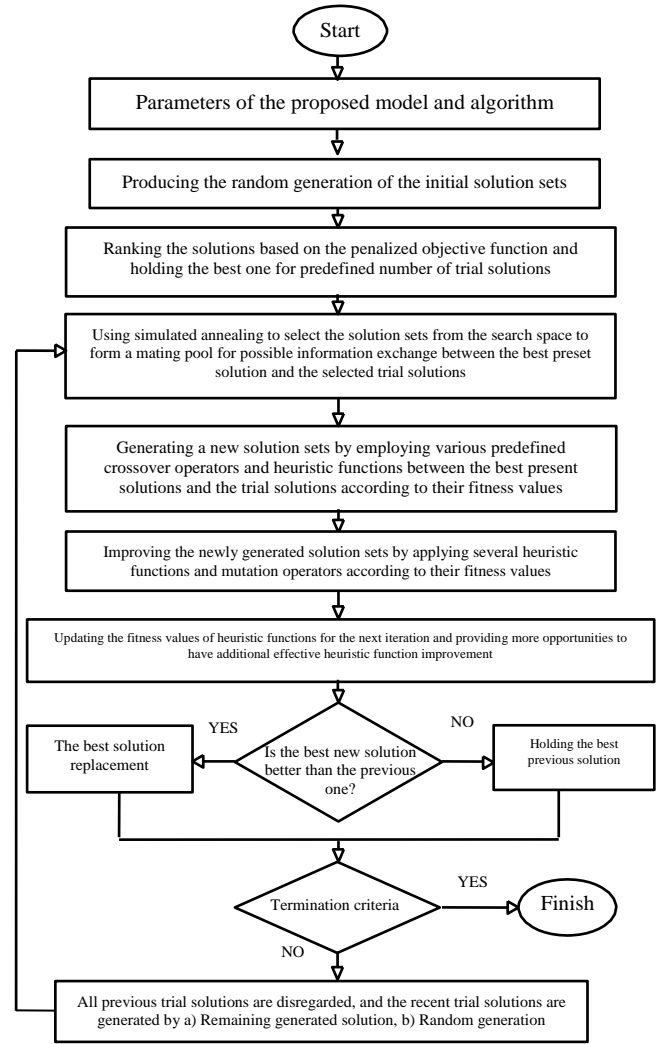

**Figure 1.** HBMO Algorithm Process for Solving the Problem [21]

### 4.2. Optimal PMU Placement Results Considering ZIBs and PMU Loss in System

Recently, power systems faced with high impact low frequency events which can affect the power system operation, monitoring and consequently cause failures which affect the system resilience [23]. For having a resilient PMU placement, in this paper PMU loss, known as measurement loss, is considered. The simulation results in Table 4 show that if the operator intend to install PMUs based on budget, for IEEE 14 bus test system 7 PMUs are required for having 100% observability even during the PMU loss scenario. Moreover, IEEE 57 and IEEE 118 bus test systems require 25 and 61 PMUs to be installed for having full observability.

### 4.3. Optimal PMU placement Results Considering Line Outages and ZIBs

In this section, optimal PMU placement results considering ZIBs, and line outages are presented according to Table 5. As shown in Table 5, 7, 19, and 53 PMUs are required for having 100% observable system in the case of any line outages. Compared to line outage scenario, it can be concluded that in the case of any PMU loss, more PMUs are required for observing the system. This can also be seen in (3) and (4) that for PMU loss the minimum required PMUs are two rather than one.

**Table 2.** Optimal PMU Placement Results without Considering ZIBs

| Test Network | Number of ZIB | PMU Locations |
|---|---|---|
| IEEE 14 bus | 4 | 2, 6, 7, 9 |
| IEEE 57 bus | 16 | 4, 7, 11, 21, 22, 24, 26, 34, 36, 37, 39, 40, 45, 46, 48 |
| IEEE 118 bus | 32 | 2, 5, 9, 11, 15, 17, 23, 25, 29, 34, 37, 41, 45, 49, 53, 56, 62, 66, 68, 70, 71, 75, 77, 80, 85, 86, 91, 94, 102, 105, 110, 114 |

**Table 3.** Optimal PMU Placement Results with Considering ZIBs

| Test Network | Number of ZIB | PMU Locations |
|---|---|---|
| IEEE 14 bus | 3 | 2, 6, 9 |
| IEEE 57 bus | 11 | 1, 6, 13, 19, 25, 29, 32, 38, 41, 51, 54, 46, 48 |
| IEEE 118 bus | 27 | 3, 8, 11, 12, 19, 22, 27, 31, 32, 34, 37, 40, 45, 49, 53, 56, 62, 75, 77, 80, 85, 86, 90, 94, 101, 105, 110 |

**Table 4.** Optimal PMU Placement Results Considering PMU Loss and ZIBs

| Test Network | Number of ZIB | PMU Locations |
|---|---|---|
| IEEE 14 bus | 7 | 1, 2, 4, 6, 9, 10, 13 |
| IEEE 57 bus | 25 | 1, 3, 4, 6, 9, 10, 12, 13, 15, 18, 20, 25, 27, 29, 30, 32, 33, 37, 39, 41, 49, 50, 53, 54, 56 |
| IEEE 118 bus | 61 | 1, 3, 7, 8, 9, 11, 12, 15, 17, 19, 21, 22, 23, 24, 27, 29, 31, 32, 34, 35, 40, 42, 44, 45, 46, 49, 51, 52, 54, 56, 57, 59, 62, 66, 68, 70, 71, 75, 76, 77, 78, 80, 83, 85, 86, 87, 89, 91, 92, 94, 96, 100, 101, 105, 106, 108, 110, 111, 112, 115, 117 |

**Table 5.** Optimal PMU placement Results Considering Line outages and ZIBs

| Test Network | Number of ZIB | PMU Locations |
|---|---|---|
| IEEE 14 bus | 7 | 2, 4, 6, 7, 9, 10, 13 |
| IEEE 57 bus | 19 | 1, 3, 6, 12, 14, 19, 21, 27, 29, 30, 32, 33, 41, 44, 49, 51, 53, 55, 56 |
| IEEE 118 bus | 53 | 1, 7, 10, 11, 13, 15, 17, 19, 21, 23, 24, 25, 27, 29, 32, 34, 35, 40, 42, 44, 46, 49, 51, 53, 56, 58, 59, 63, 69, 70, 73, 75, 76, 78, 80, 83, 85, 87, 89, 91, 92, 94, 96, 100, 102, 105, 106, 109, 111, 112, 115, 116, 117, 118 |

## 5. Conclusion

In this paper, a new optimization algorithm with the name of honey bee mating optimization (HBMO) was applied on PMU placement problem. The problem was solved based considering ZIBs, line outages, and PMU losses. Simulation results demonstrate the effectiveness of the proposed model by testing on different IEEE test systems such as 14-, 57-, 118 bus test systems. The future work will be adding channel limitation constraints on PMUs and measurement noises on this pivotal topic.

## 6. Appendix

In this paper, three IEEE test systems with 14, 57, and 118 buses are considered. The diagram of every test system is provided for better understanding about the locations of PMUs.

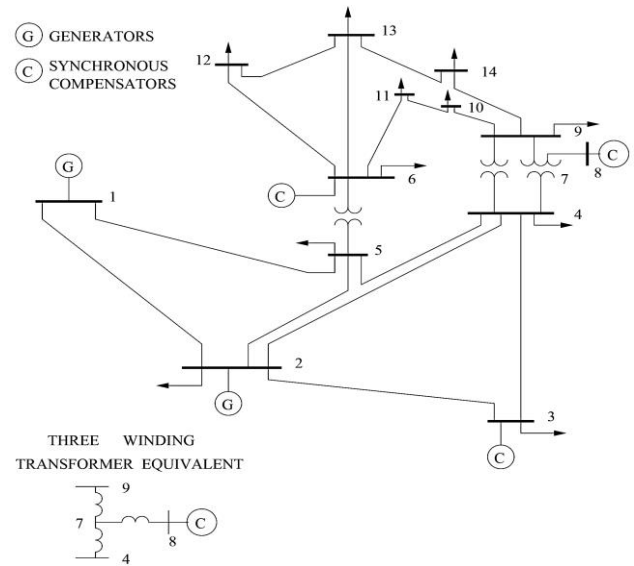

**Figure 2.** IEEE 14 bus test system

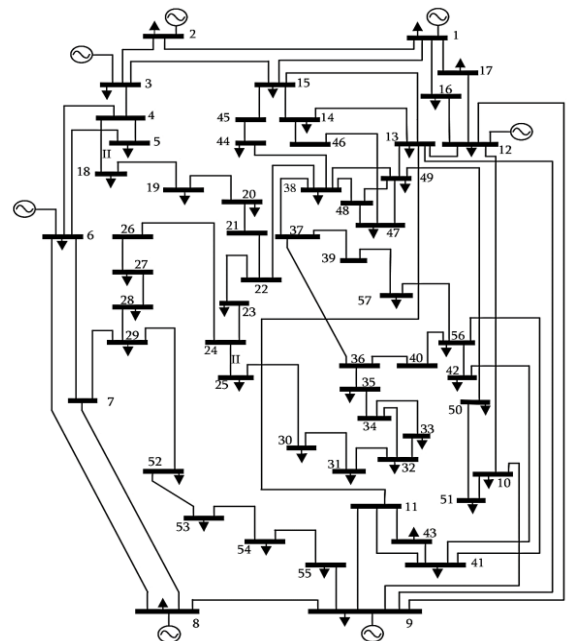

**Figure 3.** IEEE 57 bus test system

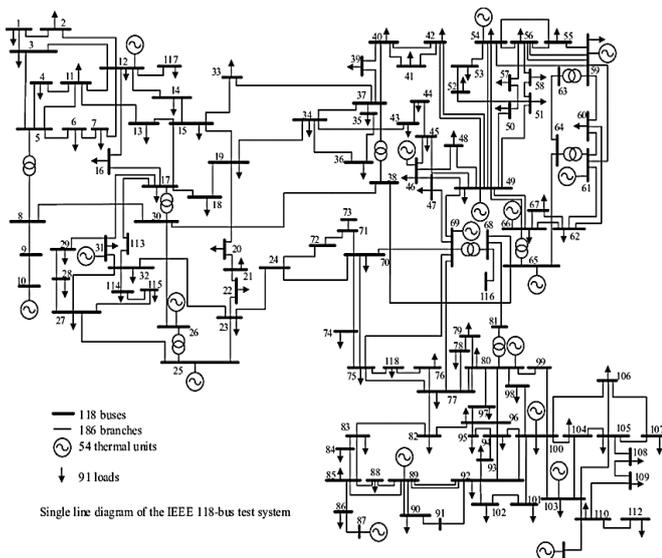

**Figure 4.** IEEE 118 bus test system